\documentclass[nocite,figures,nobm]{epl}

\usepackage{amsmath,amsfonts}
              
\title{Lorenz deterministic diffusion}
\author{R. Festa\inst{1} \and A. Mazzino\inst{2,1} \and 
        D. Vincenzi\inst{3,1}
       \thanks{E-mail: \email{vincenzi@obs-nice.fr}}
        }
\institute{
            \inst{1}INFM--Dipartimento di Fisica, 
                    Universit\`a di Genova, I--16146 Genova, Italy\\
            \inst{2}ISAC/CNR - Sezione di Lecce - Strada provinciale 
               Lecce-Monteroni km 1.2, 73100 Lecce, Italy\\
            \inst{3}CNRS, Observatoire de la C\^ote d'Azur, B.P. 4229,
06304 Nice Cedex 4, France}
\shortauthor{R. Festa \etal}

\begin{document}

\maketitle

\begin{abstract}
The Lorenz 1963 dynamical system is known to reduce in the 
steady  state to a one-dimensional
motion of a classical particle subjected to viscous damping in a past 
history-dependent potential field. If the potential field is substituted
by a periodic function of the position, the resulting system
shows a rich dynamics where (standard) diffusive
behaviours, ballistc motions and trapping take place by varying the
model control parameters. This system permits to highlight the
intimate relation
between chaos and long-time deterministic diffusion.
\end{abstract}

\section{Introduction}
Chaotic dynamical systems are known to exhibit typical random processes
behaviour due to their strong sensitivity to initial conditions.
Deterministic diffusion arises from the chaotic motion of systems
whose dynamics is specified, and it should be distinguished
from noise induced diffusion where the evolution is governed by 
probabilistic laws.
Diffusive (standard and anomalous)
behaviours have been observed in periodic chaotic maps (see, 
{\it e.g.}, refs.~\cite{Klages,Weibert} and references therein)
and in continuous time dynamical systems~\cite{Gaspard,Vulpio}.
The analysis of deterministic diffusion is relevant for the study 
of non-equilibrium processes in statistical physics. The major aim is to understand the relationship between the deterministic microscopic dynamics
of a system and its stochastic macroscopic description (think, for
example, at the connection between Lyapunov exponents, 
Kolmogorov-Sinai entropy 
and macroscopic transport coefficients, 
firstly highlighted by Gaspard and Nicolis~\cite{Nicolis}).

In the present paper we introduce a model of deterministic diffusion
in a one-dimensional lattice, suggested by a 
classical mechanics interpretation of the celebrated
Lorenz 1963 system~\cite{Lorenz}.
The steady state chaotic dynamics of the Lorenz
system
can indeed be recasted as the one-dimensional motion of a classical particle 
subjected to viscous damping in a past history-dependent potential 
field (see refs.~\cite{Festa-EPL,Festa-PRE}, and  ref.~\cite{Takeyama} for an
earlier preliminar analysis).

We shortly recall that the (scaled) Lorenz dynamical system is given by
\begin{equation}
\label{Lorenz scalato}
\left\{
\begin{array}{l}
\dot{x} = \sigma(y-x) \\
\dot{y} = -y+x+(r-1)(1-z)x\\
\dot{z} = b\,(xy-z)
\end{array}
\right.
\end{equation}
with $r>1$. For $1<r<r_{\text{c}}$, where
$r_{\text{c}} =\sigma(\sigma+b+3)/(\sigma-b-1)$,
three fixed points exist: $(0,0,0)$ (unstable)
and $(\pm 1, \pm 1 ,1)$ (stable).
For $r>r_{\text{c}}$ all fixed points are unstable, and  
the Lorenz system can exhibit either periodic or chaotic behaviour on
a strange attractor set (see, {\it e.g.}, ref. \cite{Sparrow} for
a comprehensive exposition on the subject matter).
\\
In the steady state, 
the system \eqref{Lorenz scalato} can be reduced to
a one-dimensional integro-differential equation 
for the $x$ coordinate \cite{Takeyama, Festa-PRE}
\begin{equation}
\ddot{x}+
\eta \dot{x}+(x^{2}-1)x=-\alpha [x^{2}-1]_
\beta x\, ,
\label{eq.Lorenz}
\end{equation}
where  $\alpha=(2\sigma/b)-1$, $
\beta= [2b/(r-1)]^{1/2}$, $\eta =(\sigma + 1)/[(r-1)b/2]^{1/2}$,
and the time is scaled by a factor 
$[(r-1)b/2]^{1/2}$ with respect to the time coordinate
in eqs.~\eqref{Lorenz scalato}.
The square brackets in eq.~\eqref{eq.Lorenz} indicate the {\em 
exponentially
vanishing memory}, which is defined, for any suitable time 
function 
$f(t)$, by
\begin{displaymath}
\label{exponential memory}
[f]_k(t)\equiv
k\int_0^\infty \upd s \,e^{-ks}f(t-s)\, .
\end{displaymath}
According to eq.~\eqref{eq.Lorenz}, the Lorenz system chaotic dynamics
corresponds to a one dimensional motion in a 
constant-in-time quartic potential
$U(x)=(x^2-1)^2/4$. In the presence of friction
($\eta\ne 0$) the motion is sustained by a time dependent
memory term, which takes into account the system past evolution.

Although eq.~\eqref{eq.Lorenz} has been deduced from the 
Lorenz system~\eqref{Lorenz scalato}, it can be generalized to a wider
class of equations showing similar dynamical properties~\cite{Festa-PRE}.
Indeed, it can be usefully recasted in the form 
\begin{equation}
\label{eq. generalizzata}
\ddot{x}+
 \eta \dot{x}+ 
\{ q(x)+\alpha
\left[q(x)\right]_\beta \}\Phi'(x)=0
\end{equation}
where the prime indicates the derivative with respect to $x$. 
Equation~\eqref{eq.Lorenz} is obtained for 
$\Phi(x)=x^2/2$ and $q(x)=x^2-1$. 
The generalized equation~\eqref{eq. generalizzata} can be
regarded as the description of the motion of a unit mass particle subjected
to a viscous force $-\eta\dot{x}$ and interacting with a potential
field $\Phi(x)$ through a dynamically varying ``charge''
$q_t (x)=q(x)+\alpha [q(x)]_\beta$. 
This charge depends both
on the instantaneous particle position $x(t)$ 
and  on the past history $\{x(t-s)\,|\, 0\le s<\infty\}$.
It is just the coupling of $[q(x)]_\beta$ with the fixed potential field 
$\Phi(x)$ the origin of an endogenous forcing term which can sustain the
motion even in the presence of friction: the chaotic 
behaviour can actually arise from the synergy
between this term and the viscosity.  \\Moreover, one can easily verify that 
eq.~\eqref{eq. generalizzata} corresponds to 
the {\em generalized Lorenz system}
\begin{equation}
\label{sist.gen.}
\left\{ \begin{array}{l}
\dot{x}=\sigma(y-x) 
\\
\dot{y}=-y+x+(r-1)(1-z)\,\Phi'(x)
\\
\dot{z}=-bz+b[\frac{1}{2}
q'(x)(y-x) +q(x)+1]\, .
\end{array}
\right.
\end{equation}
The specific Lorenz model can thus be viewed as  
singled out from a quite general class of dynamical systems
which can exhibit chaotic behaviour, their common essential
property being an exponentially vanishing memory 
effect together with a viscous damping.

In our previous works~\cite{Festa-EPL,Festa-PRE}
 the main chaotic dynamical features of the
original Lorenz system have been investigated  
through the analysis of the piecewise
linear system corresponding to the choice 
$\Phi (x)=|x|$ and 
$q(x)=|x|-1$. 
We have thus obtained the piecewise linear Lorenz-like equation 
\begin{equation}
\label{eq.lin.}
\ddot{x}+\eta \dot{x}+ 
\{|x|-1 
+\alpha \left[|x|-1\right]_{\beta}\}
\mbox{sgn}(x)=0
\end{equation}
which has the great advantage of being exactly solvable on each side
of $x=0$. 

\section{The Lorenz diffusion}

\begin{figure}
\onefigure[height=3.5cm]{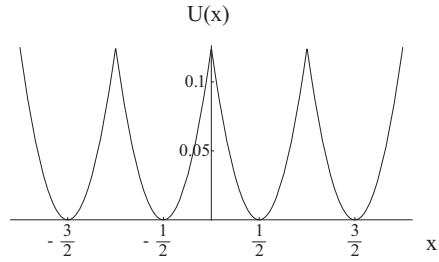}
\caption{Periodic lattice of truncated parabolae. The system evolves
in the periodic constant-in-time potential $U(x)$ subjected to a viscous
damping and to a memory effect on the past motion.}
\label{fig:reticolo}
\end{figure}

If in eq.~(\ref{eq. generalizzata}) one substitutes the quantities
$q(x)$ and $\Phi'(x)$ with $x$-periodic functions, the chaotic jumps between
the two infinite wells 
of the original quartic potential $U(x)$ correspond to chaotic
jumps among near cells. The result is a deterministic diffusion
in an infinite lattice, induced by a Lorenz-like chaotic dynamics.
Equation~(\ref{eq. generalizzata})  will be thus called the
{\it Lorenz diffusion equation}.
In order to use as far as possible analytical tools,
we shall consider the unit wavelength periodic potential
$
U(x) =\frac{1}{2}{\left(\{x\}-\frac{1}{2}\right)}^2$ (corresponding to $q(x)=\{ x\}-\frac{1}{2}$
and $\Phi(x)=\{ x\}$),
where $\{x\}$ indicates the fractionary part of $x$. This potential field
obviously consists of a lattice of truncated parabolae (see fig.~\ref{fig:reticolo}).
By simple substitution
one easily derives the equation
\begin{equation}
\label{per.eq.}
\ddot{x}+\eta \dot{x}+ 
\{x\}-\textstyle\frac{1}{2} +\alpha [\{x\}-\frac{1}{2}]_{\beta} \,=0\,
\end{equation}
where $\eta$ denotes the 
friction coefficient, $\alpha$ is the memory amplitude and $\beta$ is
related to the inertia whereby the system keeps memory 
of the past evolution.
Note that $\eta$ and $\beta$ play a symmetrical
role in the dynamics: the solution of eq.~\eqref{per.eq.} is 
indeed left
invariant if one changes $\beta$ with $\eta$, while keeping 
$\beta(1+\alpha)$ constant.

Inside each potential cell
eq.~\eqref{per.eq.} can be recasted in a third-order linear differential form.
Indeed,
by applying the operator $(d/dt + \beta)$ to each side
of eq.~\eqref{per.eq.}, 
one obtains (for $x\ne n$)  
\begin{equation}
\label{lin.cell.1}
\frac{\upd^3 x}{\upd t^3}+(\beta+\eta)\frac{\upd^2 x}{\upd t^2}+
(1+\beta\eta)\frac{\upd x}{\upd t}+\beta(1+\alpha)(x-n-\frac{1}{2})=0\, .
\end{equation}
It is worth observing that 
the nonlinearity of the original model is simply reduced to
a change of sign of the forcing term $\beta(1+\alpha)(\{x\}-\frac{1}{2})$ 
when $x$ crosses the cell boundaries (in our case the integer values).
As we will see, chaotic dynamics essentially results from the unpredictability
of the crossing times.\\
Partial solutions of the third order nonlinear differential
equation can be easily calculated inside each open interval
$(n,n+1)$. To obtain a global solution,
such partial solutions should be matched 
at $x=n$ by  
assuming that the position $x$, 
the velocity $\dot{x}$
and the memory $[\,|x|\,]_\beta$ are
continuous, whereas the acceleration $\ddot{x}$ turns out to be undefined.  
However, it is easily shown
that each pair of acceleration values ``immediately''
before and after the crossing times are related by $
\ddot{x}^{(+)} 
-\ddot{x}^{(-)}
=\text{sgn}(\dot{x})
$.\\
The fixed points of eq.~\eqref{per.eq.} are
of course $x=n+\frac{1}{2}$, $n\in\mathbb{Z}$, and their local stability 
depends on the roots of the characteristic equation
\begin{equation}
\label{l}  
\lambda ^3 + (\beta + \eta)\lambda^2 +
(1 + \beta \eta)\lambda + \beta (1 + \alpha) = 0\, .
\end{equation}
All the fixed points are unstable when $\alpha$ is 
larger than the critical value 
$\alpha_{\text{c}}=\beta^{-1}(1+\beta \eta)(\beta + \eta)-1$.
In this case eq.~\eqref{l}
has one real negative root $(-\lambda_0<0)$ and a complex
conjugate pair of roots with positive real part 
$\lambda_\pm=\lambda\pm i\omega$ $(\lambda>0)$.
For $\alpha>\alpha_{\text{c}}$,
the partial solution in the generic open interval
$(n,n+1)$ can be finally written in the explicit form
\begin{equation}
\label{sol.gen.}
x(t)= e^{\lambda t} \left(\, C_1 \cos (\omega t) \,+\,
C_2 \sin (\omega t)\, \right)\:+\: 
C_3 \, e^{ -\lambda _0 t}+n+1/2
\end{equation}
where the constants $C_1$, $C_2$, $C_3$ are linearly related
to the (cell by cell) initial conditions.
The
motion inside each cell consists
of an amplified oscillation around a central point which translates
towards the center of the cell. A change of cell yields
a discontinuous variation of the acceleration $\ddot x$ and consequently
of the coefficients $C_1$, $C_2$, $C_3$.

\begin{figure}
\onefigure[height=5.3cm]{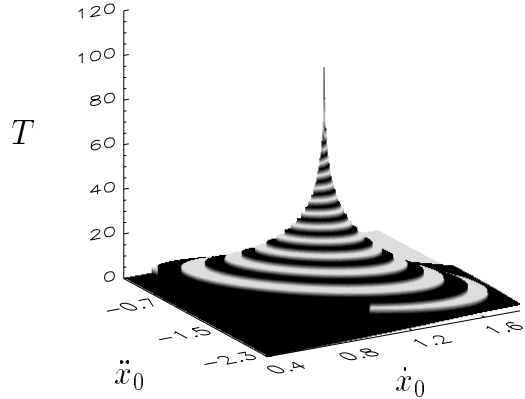}
\caption{Dependence of the exiting time from the elementary cell
on the initial conditions $\dot{x}_0$,
$\ddot{x}_0$ for $\alpha=6.50$, $\beta=0.19$, $\eta=0.78$. 
The graph has been obtained via numerical solution of eq.~\eqref{eq:T}.
Backward ($x=n$)
and forward ($x=n+1$) exiting cases have been distinguished by different
colors. The structure of the graph can be guessed through the analysis
of the contour lines of the function $T=T(\dot x _0, \ddot x_0)$ 
(see ref.~\cite{Festa-PRE} for further details).}
\label{fig:T}
\end{figure}

Suppose that, at a given time, say $t=0$, the particle enters 
the $n$th cell
at its left boundary with positive velocity (the reverse case can be 
symmetrically analyzed). In this case $C_3=-(C_1+\frac{1}{2})$.
The question is now on whether the particle leaves the cell 
either from the left side (i.e. $x=n$)
or from the right side of the cell (i.e. $x=n+1$).
Assigned the model
parameters $\lambda_0$, $\lambda$ and $\omega$, the minimum positive
time $T$ such that
\begin{equation}
\label{eq:T}
\left|e^{\lambda T} \left(\, C_1 \cos (\omega T) \,+\,
C_2 \sin (\omega T)\, \right)\, 
-\, (C_1+1/2)\, e^{ -\lambda _0 T}\right|=1/2
\end{equation}
depends of course on $C_1,C_2$, and therefore on 
the initial
conditions $\dot{x}_0,\ddot{x}_0$.  Unfortunately, the direct problem
is transcendent. Moreover, as shown in fig.~\ref{fig:T}, its solution
is strongly sensitive to the initial conditions.
This fact is a 
direct consequence of the crossing time definition:
$T\/$ is indeed determined by the intersection of an amplified
oscillation and a decreasing exponential. A small change in the
initial conditions may thus
cause a 
discontinuous variation of the crossing time.
This is the very origin of the system chaotic dynamics
which suggests a stochastic treatment of the Lorenz diffusion equation.

Despite the fact we have a three-dimensional space of parameters
to investigate the model behaviors, the interesting region is
actually a limited portion. This easily follows from the 
following
simple considerations.
 For $\eta$ large enough in eq.~\eqref{per.eq.}, 
the motion rapidly stops in one of the lattice
fixed points. 
In order to have non trivial solutions, the viscous coefficient 
must be smaller than a maximum value which can be explicitly
derived from the
condition $\alpha>\alpha_{\mbox{c}}$
$$
\eta_{\text{max}}=\frac{1}{2\beta}\left[
\sqrt{{(1+\beta^2)}^2+4\alpha\beta^2}-1-\beta^2\right]\, .
$$
In the opposite limit, if $\eta$ is too small, the friction term is negligible
with respect to the memory term, and the resulting motion is 
``ballistic'', {\it i.e.}, $\left< [x(t)-x(0)]^2\right>
\sim t^2$\footnote{
Due to the deterministic nature of the system, 
averages 
should be intended over the initial conditions.
In our numerical simulations we have typically chosen
the initial conditions to be uniformly
distributed over some real interval, and we have 
naturally assumed that
the diffusion coefficient is independent of the choice of the initial 
ensemble.}. Analogous considerations
can be repeated for $\alpha$ and $\beta$. 
Inside this region of parameters the Lorenz
diffusion equation generates a wide variety of behaviours: as we will see,
the observed regimens are strongly 
sensitive to the control parameters, and furthermore this dependence is 
often in contrast with the intuitive meaning of $\alpha$, $\beta$,
$\eta$.

\begin{figure}
\twoimages[height=4cm]{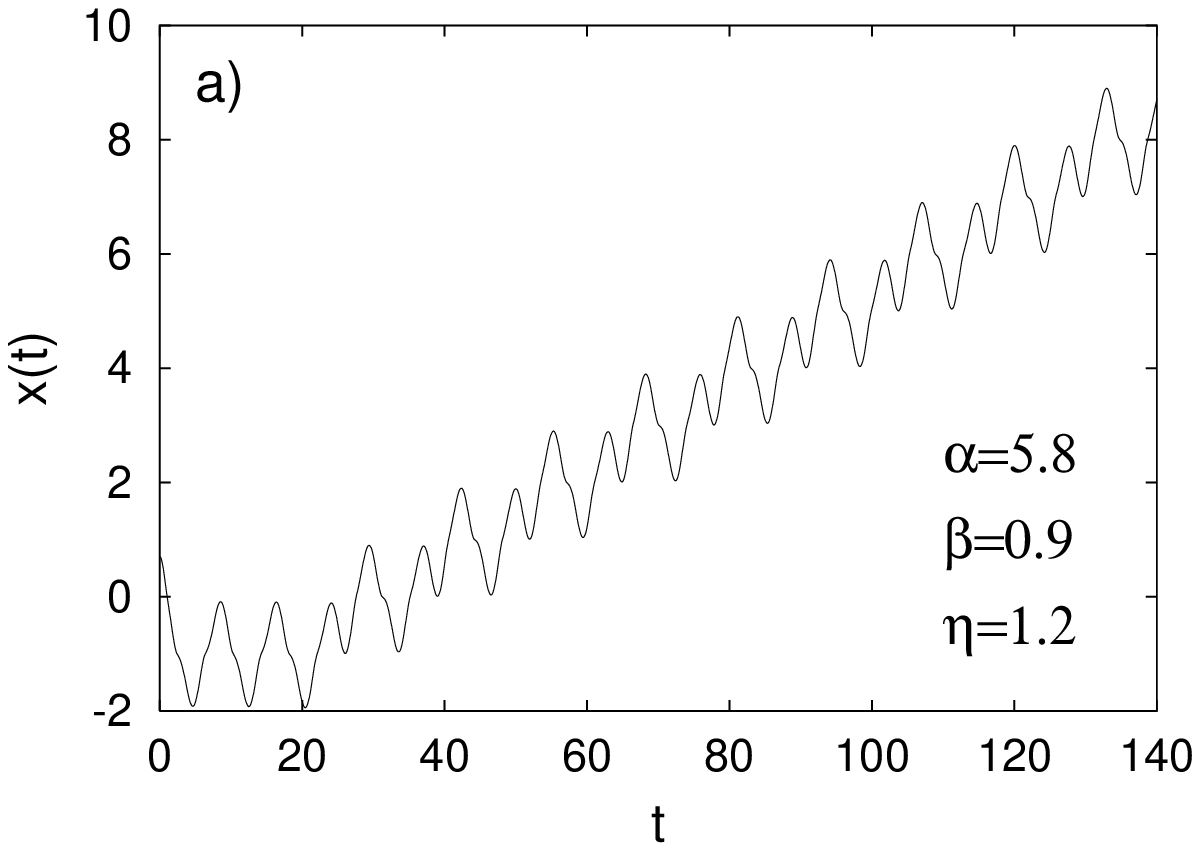}{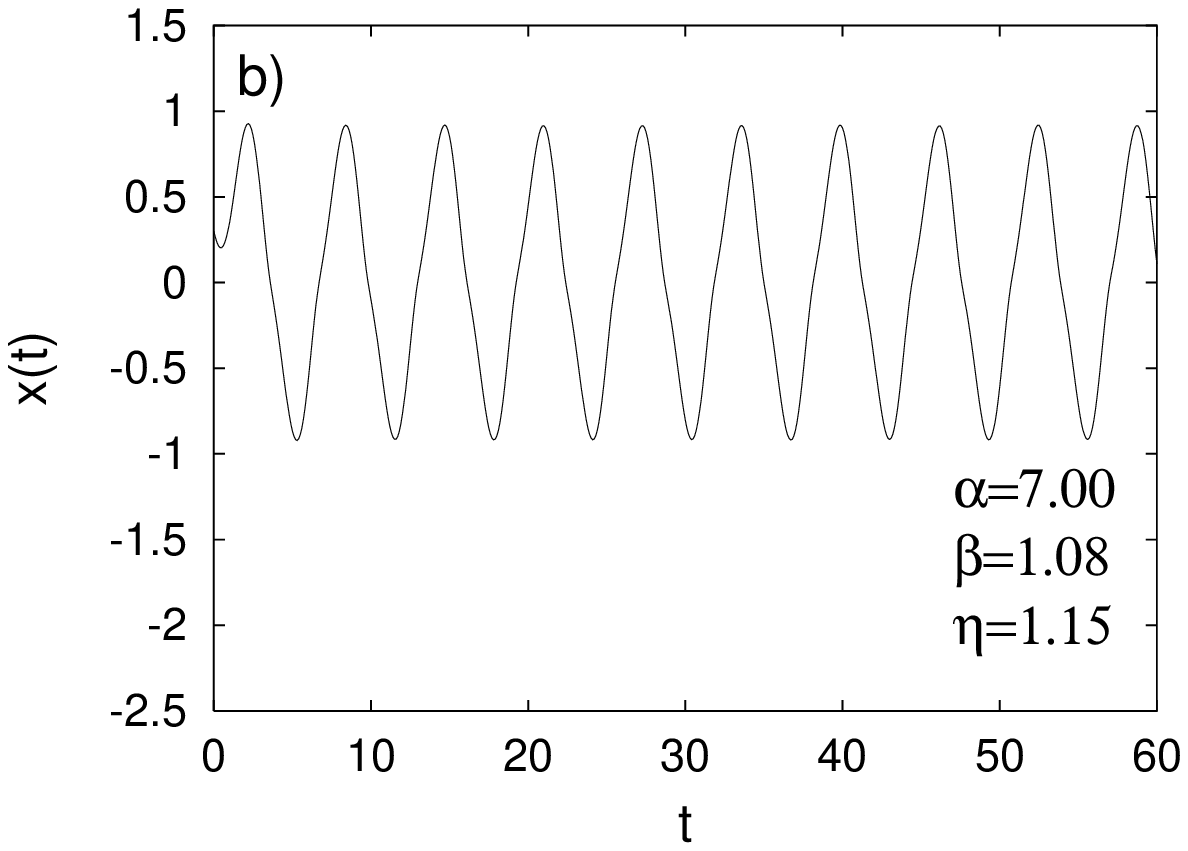}
\twoimages[height=4cm]{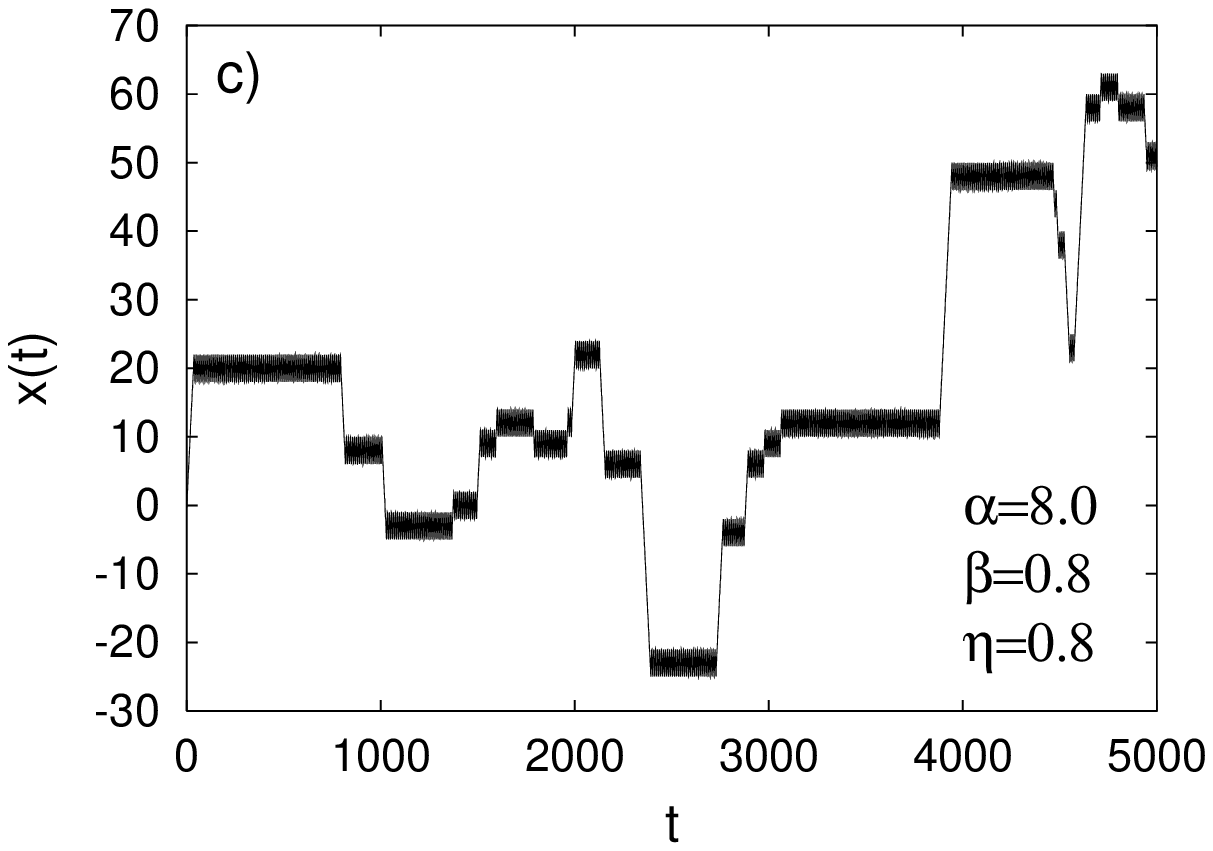}{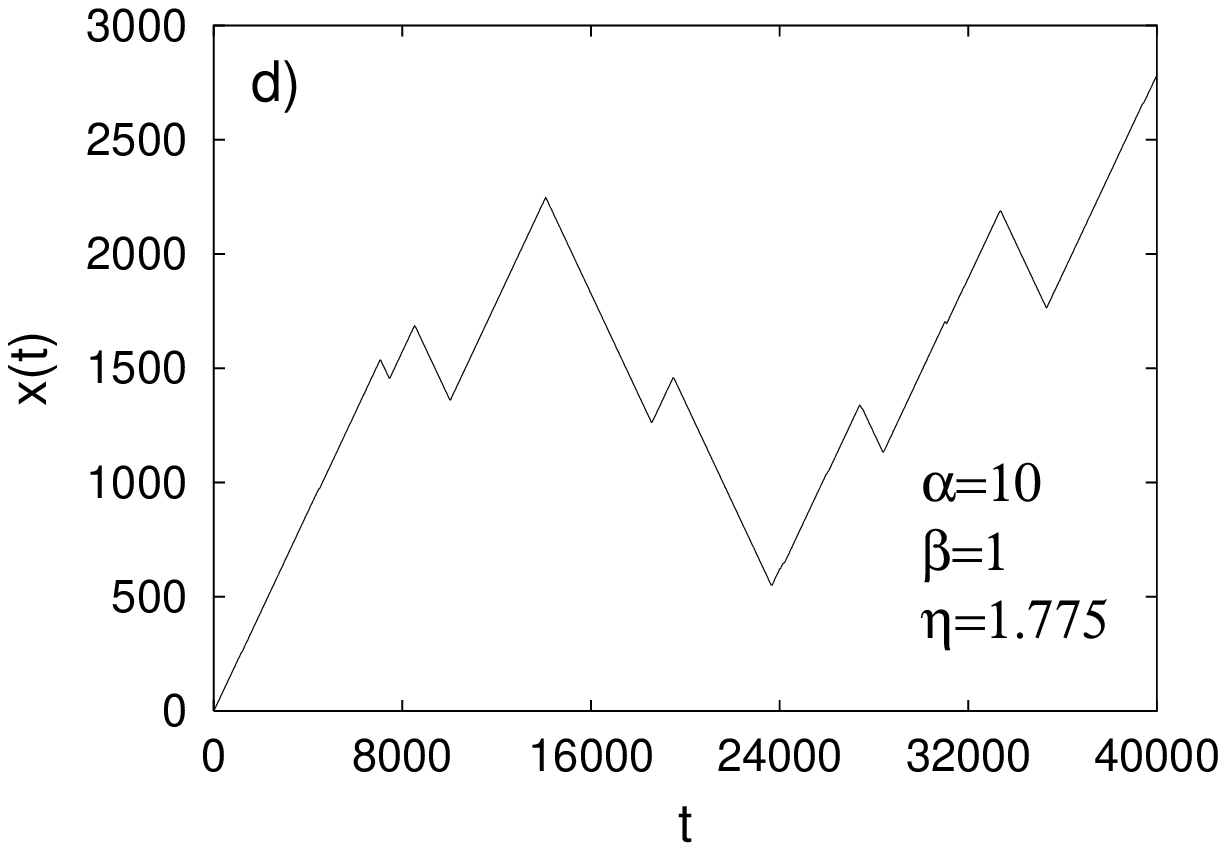}
\twoimages[height=4cm]{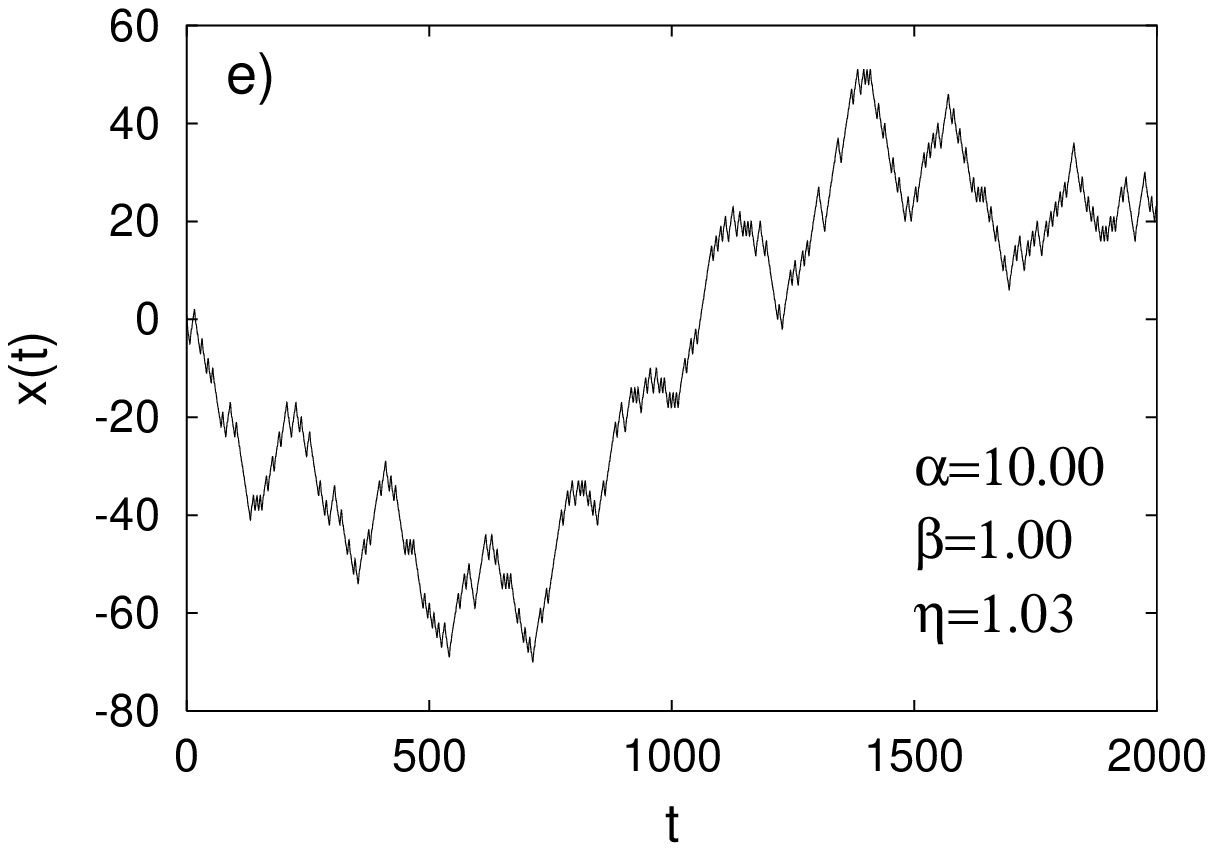}{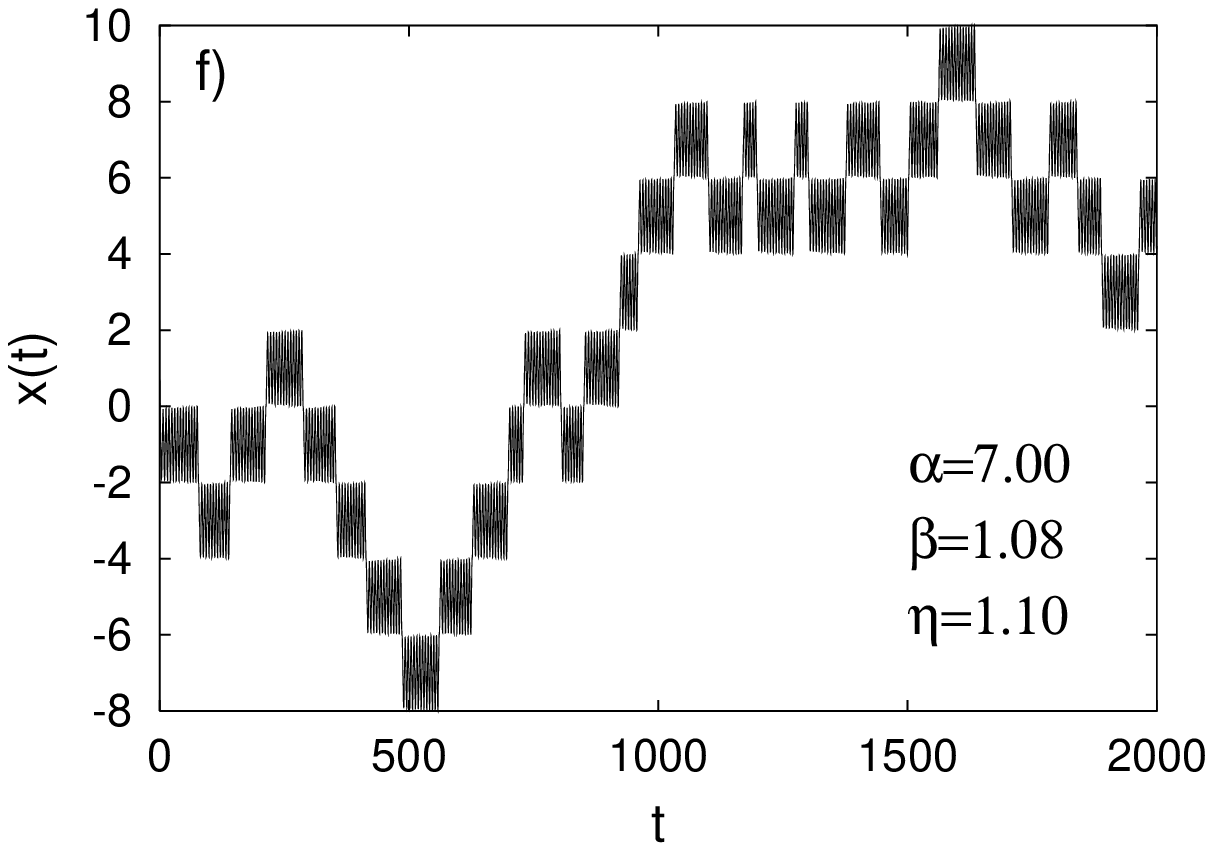}
\caption{{\it a}) A typical ballistic motion with a more
complex structure than a simple crossing of adjacent cells. Ballistic
motion generally corresponds to a periodic behaviour in the cell crossing;
{\it b}) trapping in two cells; 
{\it c})-{\it f}) different kinds of trajectories generating diffusive motions.}
\label{regimi}
\end{figure}

In fig.~\ref{regimi} a few numerical simulation of
eq.~\eqref{per.eq.} are shown, corresponding to different values of the parameters. These examples suggest that the variety of motion regimens
ranges from  ``ballistic'' ones  to clearly ``diffusive'', and
even ``trapped'' in one cell or in groups of nearby cells.

\begin{figure}
\twoimages[height=5cm]{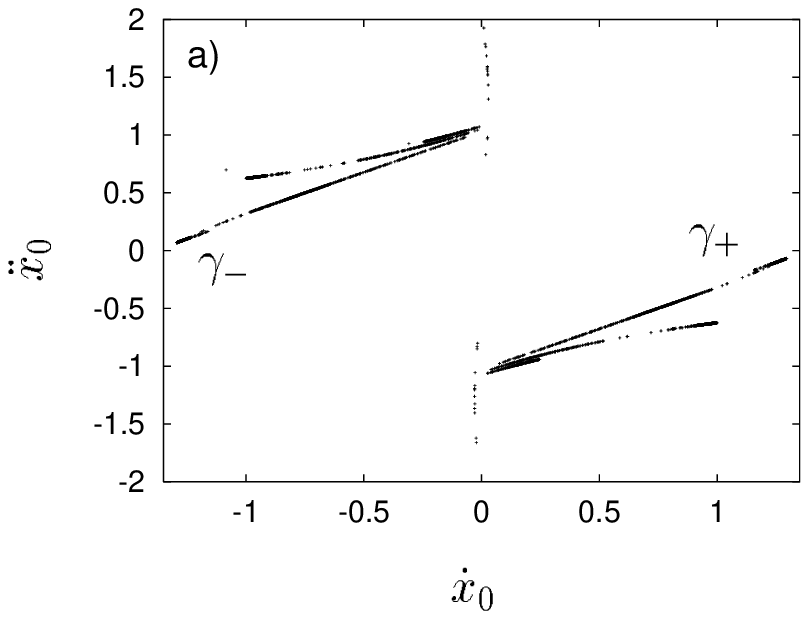}{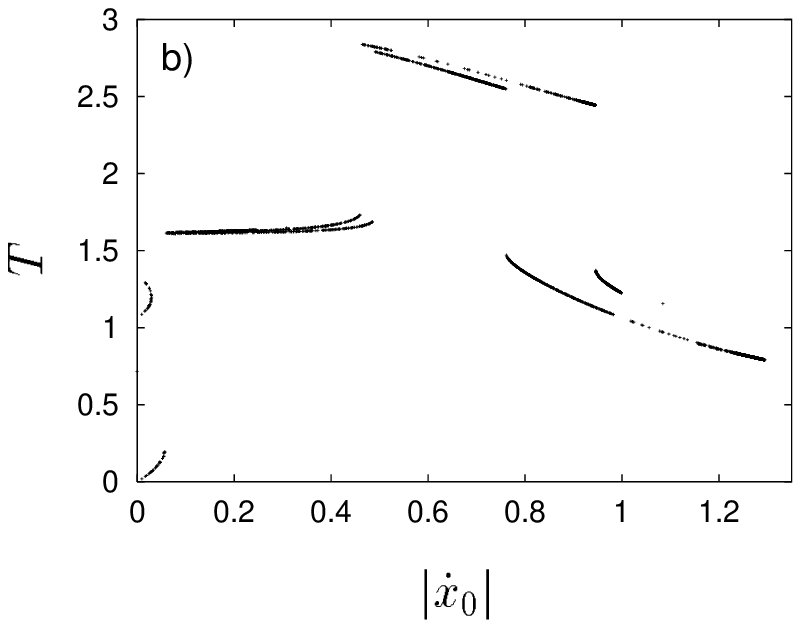}
\caption{{\it a}) A typical attracting set for the $n$th cell entering
conditions $\dot{x}_0$, $\ddot{x}_0$. The corresponding trajectory
is shown in fig.~\ref{regimi}{\it e}). It should be noted the presence
of the
straight lines defined by eq.~\eqref{retta}, 
here denoted by $\gamma_+$ and $\gamma_-$;
{\it b}) attracting set in the plane $(T,\dot{x}_0)$. The step-like
structure of the set derives from the superposition of 
the points in {\it a}) on a graph like that of fig.~\ref{fig:T}.}
\label{fig:mappe}
\end{figure}

The analysis of the motion in the elementary cell shows that the system
rapidly reaches a steady-state. In the diffusive regimens the points corresponding to subsequent cell entering conditions $\dot x_0$, $\ddot x_0$ are
quickly attracted on a particular locus of the plane (fig.~\ref{fig:mappe}).
In the general situation this attracting set does not define a univocal map
between the entering velocity and acceleration. However,
it is not difficult to see that, for large $T$, $\dot{x}_0$ and $\ddot{x}_0$
satisfy the piecewise linear relation~\cite{Festa-PRE}  
\begin{equation}
\label{retta}
\ddot{x}_0-2\lambda\dot{x}_0+\frac{1}{2}\left(2-
\lambda^2-\omega^2\right)\text{sgn}(\dot{x_0})=0\, .
\end{equation}
If the diffusive regimen admits large enough times of permanence,
the couple of straight lines defined by eq.~\eqref{retta} therefore
belongs to the entering condition attracting set (fig.~\ref{fig:mappe}). 
Moreover, in 
the limiting case of very large $\lambda_0$ the attracting set exactly
reduces to such lines, and in the steady state 
eq.~\eqref{retta} is satisfied by all entering
conditions~\cite{Festa-PRE}. 

\begin{figure}
\twoimages[height=4cm]{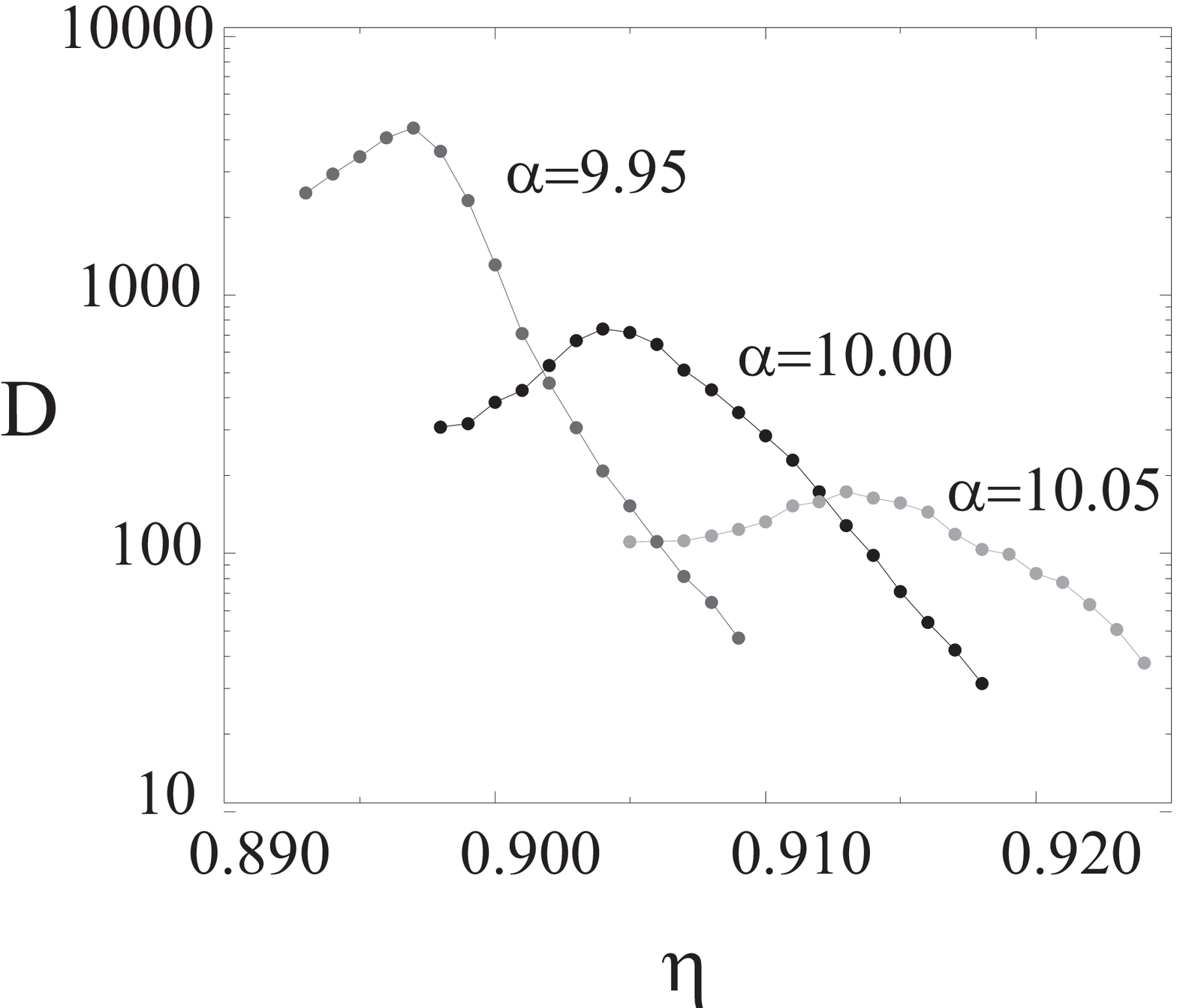}{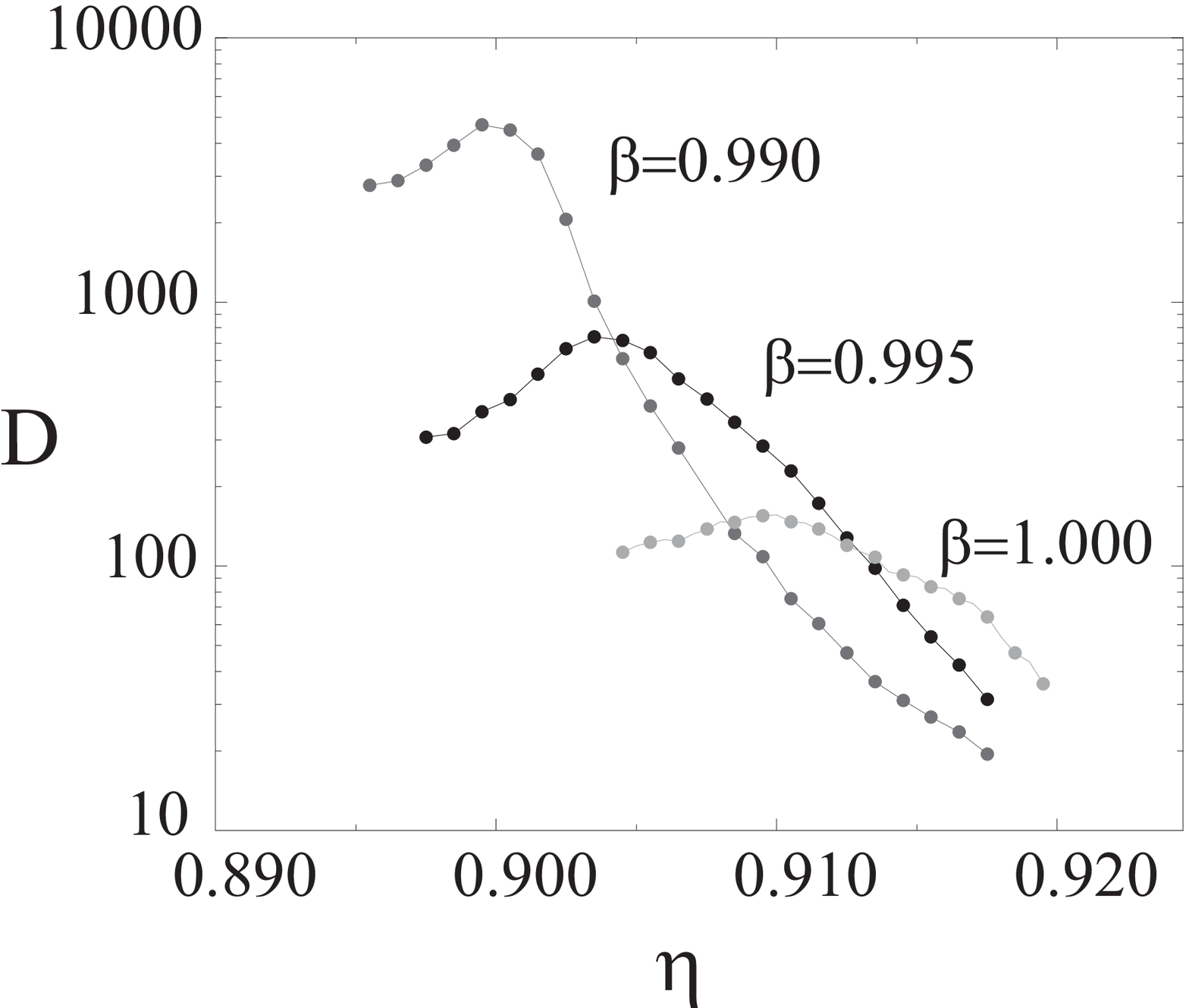}
\caption{Dependence 
of the diffusion coefficient $D$ on the viscosity parameter $\eta$
for different values of $\alpha$ and $\beta$.}
\label{rison}
\end{figure}

The macroscopic evolution of the system can be statistically analyzed 
by introducing the diffusion coefficient 
$D\equiv\lim_{t\to\infty}\left<[x(t)-x(0)]^2\right>/(2t)\,.$
Diffusive regimens are identified by a finite value of $D$. As shown in 
fig.~\ref{regimi}, different values of the control parameters 
can lead to qualitatively very different diffusive motions.
What is surprising is that the change of regimen is very 
irregular and abrupt with $D$, and sometimes it does not
reflect our intuition on the
role played by the parameters in eq.~\eqref{per.eq.}.
For instance, $D$ shows a non-monotonic dependence on the viscosity
$\eta$ (fig.~\ref{rison}), and it is possible to observe transitions 
from trapped (or diffusive) motion to ballistic ones even
by increasing the viscosity itself (table~\ref{table:D}).

Despite the somehow strange aspect of some system trajectories,
it should be stressed that no anomalies
({\it i.e.} superdiffusive transport)
have been encountered in the statistical
analysis of diffusion, the distribution of long flies
having always exponential tails.
%
%
\begin{table}
\caption{
Different long-time  behaviors as a function of $\eta$, for fixed
$\alpha$ and $\beta$. Note the nontrivial, highly structured
dependence of the diffusion coefficient on the control parameter.
}
\label{table:D}
\begin{center}
\begin{largetabular}{|c|c|c|c|}
\hline
$\eta$ & REGIMEN&$\eta$&REGIMEN\\
\hline
$<0.850$ & ballistic&1.425-1.725 & trapped\\
$0.850-0.875$ & trapped&1.725-1.975 & diffusive\\
$0.875-0.900$ & ballistic&1.975-2.050& trapped\\
$0.900-1.050$ & diffusive&2.050-2.325 & diffusive\\
$1.050-1.175$ & trapped&$2.325<$& trapped\\
$1.175-1.425$ & diffusive&& \\
\hline
\end{largetabular}
\end{center}
\end{table}

To conclude,
we have presented  a deterministic system which can generate
large-scale diffusive transport induced by a  Lorenz-like microscopic
chaotic dynamics. Two main problems should still be tackled.
First, the diffusion coefficient seems to have a fractal dependence
on the control parameters, as observed for other discrete 
chaotic maps~\cite{Klages}. This fact requires more investigations, 
possibly using the analytical tools developed in ref.~\cite{Festa-PRE}.
Second, it would be interesting to consider a two-dimensional
extension of the Lorenz diffusion equation, mainly to 
investigate the possible emergence of anomalous diffusion induced
by the increased spatial dimension. 

\acknowledgments

The authors would like to thank M. LA CAMERA for useful
suggestions. D. V. was supported by
grants of the University of Nice and 
of the University of Genova.
This work has been partially supported by the INFM project GEPAIGG01
and Cofin 2001, prot. 2001023848.

\end{document}